\documentclass {elsart}
\begin {document}
\begin{frontmatter}

\title{
Transverse and in-plane modification of superconductivity and
electronic structure in the quasi-two dimensional organic
conductor $\kappa $--(BEDT-TTF)$_{2}$Cu(SCN)$_{2}$ by uniaxial
stress}

\author [FSU,SNU]{E. S. Choi}
\author [FSU]{J. S. Brooks}
\author [FSU]{S. Y. Han}
\author [FSU,venezuela]{L. Balicas}
\author [FSU]{J. S. Qualls}

\address [FSU]{National High Magnetic Field Laboratory and Florida State
University, Tallahassee, FL 32310, USA}
\address [SNU] {Research Institute for Basic Sciences, Seoul National
University, Seoul 151-742, Korea}
\address [venezuela] {Instituto Venezolano de Investigaciones
Cient\'{\i}ficas, Apartado 21827, Caracas 1020A, Venezuela}

\begin{abstract}

We have employed uniaxial stress along the principal axes of the
quasi-two dimensional organic superconductor $\kappa
$--(BEDT-TTF)$_{2}$Cu(SCN)$_{2}$. The lattice anisotropy is
thereby altered, with corresponding changes in the intermolecular
transfer energies. The effect of uniaxial stress on the
superconducting transition temperature $T_{c}$ and critical field
$B_{c2}$ is found to be anisotropic.There is an indication of an
increase in $T_{c}$ and $B_{c2}$ for in-plane stress, but both
parameters decrease rapidly for transverse (inter-plane) stress.
Magnetotransport studies reveal stress-induced changes in the
Fermi surface through the observation of the Shubnikov de Haas
oscillations. The stress dependence of a resistive anomaly in the
magnetoresistance, which is associated with the critical field
$B_{c2}$, is also investigated. We discuss the experimental
findings in the context of recent phenomenological and theoretical
treatments of quasi-two dimensional systems where the anisotropic
triangular lattice Hubbard model has been used to treat
two-dimensional superconductors.
\end{abstract}

\begin{keyword}
A. Organic crystal; A. Superconductors; D. Electronic transport;
E. High pressure
\end{keyword}

\end{frontmatter}
\newpage

\section{Introduction}
The quasi-two-dimensional organic superconductor families, $\kappa
$--(BEDT-TTF)$_{2}$X (X = Cu(NCS)$_{2}$, Cu[N(CN)$_{2}$]Br,
Cu[N(CN)$_{2}$]Cl), have been widely studied both experimentally
and theoretically \cite{ishiguro}. Depending on the species of the
anion (X) and the pressure, these compounds can exhibit
ferromagnetic, antiferromagnetic, insulating, or superconducting
ground states. These "kappa-phase" materials have been the subject
of considerable recent attention due to reports of evidence for
d-wave superconductivity\cite{schrama}\cite{naka},
\cite{carrington},\cite{tomic}, as well as reports to the contrary
\cite{wosnitza2}. Fig. 1 shows the molecular stacking and the
Fermi surface of $\kappa $--(BEDT-TTF)$_{2}$Cu(SCN)$_{2}$, which
is the subject of this report. This material is a superconductor
with a transition temperature $T_c$ = 10.4 K.
 Theoretically, the
anisotropic triangular lattice Hubbard model has been used to
treat the kappa-phase ground
states\cite{kino96,kino98,kondo98,mckenzie,vojta,kondo99}. In the
triangular lattice model, each BEDT-TTF dimer is treated as a
single lattice point in the triangular lattice with four nearest
neighbors with transfer integral $t_{2}$ and two next nearest
neighbors with transfer integral $t_{1}$ (See Fig. 1a). Hence the
system is modeled as a triangular arrangement with one electron on
each vertex site. For favorable ratios of $t_{1}$ and $t_{2}$,
unconventional pairing (superconductivity) may arise from
antiferromagnetic fluctuations,  with the interaction modeled by
the on-site Coulomb repulsion $U$. The consensus of these models
is that the most likely superconducting order parameter symmetry
is d-wave. Further more, Kondo and Moriya have
predicted\cite{kondo98,kondo99} a maximum in $T_{c}$ for
particlular values of $t_{1}$ and $t_{2}$ (see Discussion
section).

The motivation for the present work has been, therefore, to apply
in-plane uniaxial stress in order to modify $t_{1}$ and $t_{2}$,
and to monitor corresponding changes in the transition
temperature, and other physical properties in a $\kappa $-phase
superconductor with high magnetic fields. The most important
results are:

1) the very weak dependence of $T_{c}$ and $B_{c2}$ on in-plane
stress, which in some cases actually increase;

2) the stress-dependent nature of the resistive anomaly in the
magnetoresistance which has been ascribed to unconventional
superconductivity;

3) through the observation of the Shubnikov de Haas oscillations,
we describe changes in the Fermi surface with stress, and
indirectly, changes in the lattice constants.

For completeness, in Tables I and II we present a summary of
previous and present results which treat $T_c$ and corresponding
changes in the Fermi surface topology of $\kappa
$--(BEDT-TTF)$_{2}$Cu(SCN)$_{2}$ salt under hydrostatic pressure,
tensile stress, thermal expansion, and uniaxial stress.
%%%%%%%%%%%%%%%%%%%%%%%%%%%%%%%%%%%%%%%%%%%%%%%%%%%%%%%%%%%%%%%%%%%
\section{Experiment}
$\kappa $--(BEDT-TTF)$_{2}$Cu(SCN)$_{2}$ single crystals were
grown by conventional electrochemical crystallization. Typical
dimensions of the samples were (for $a\times b\times c$)
$0.1\times0.6\times0.9$ mm$^{3}$ with the  $b$-$c$ plane
corresponding to the most conducting plane. Uniaxial stress was
applied by using the epoxy encapsulation method \cite{camposrsi}.
Polarized infrared (IR) reflectance in the range of 650 $-$ 3500
cm$^{-1}$ was employed to determine the $b$- and $c$-axis
orientations \cite{ugawa}. Samples were cut into two or three
pieces to use for the three orientations, $a$-, $b$-, or $c$-axis,
and  electrical contacts were made with 12.5 $\mu$m gold wires
attached with silver paste. A conventional 4-terminal AC technique
was used with a frequency of $\sim$ 17 Hz, and an ac current of 10
$\mu$A was applied along the least conducting axis ($a$-axis). The
magnetic field was applied along the a-axis, perpendicular to the
conducting b-c plane. For magnetoresistance measurements where
stress in the b-c plane was studied, a novel device was used to
transmit stress along the in-plane direction with the applied
magnetic field perpendicular to the direction of stress
\cite{brooksUni}. In all cases, stress was applied in situ, at low
temperatures, in increasing increments.

%%%%%%%%%%%%%%%%%%%%%%%%%%%%%%%%%%%%%%%%%%%%%%%%%%%%%%%%%%%%%%%%%%%
\section{Results}
\subsection{Uniaxial stress dependence of $T_c$}
 Fig. 2 shows the change of the superconducting
transition $T_c$ of $\kappa $--(BEDT-TTF)$_{2}$Cu(SCN)$_{2}$ for
uniaxial stress along the in-plane, c-axis direction.  The value
of $T_{c}$ obtained from the ambient pressure (no epoxy)
measurements is almost identical to that of the samples with epoxy
(at zero pressure) , but some of the samples showed a wider
transition width and/or finite residual resistance below $T_{c}$.
In this work we define $T_c$ from the maximum in the
resistance-temperature derivative ($dR/dT$) in the range of the
transition, as shown in the inset of Fig. 2. Hence over the range
of 8 kbar, $T_c$ decreases only by about 20\%, but the normal
state resistance decreases by a factor of six.

A summary of the uniaxial stress experiments along the $a$-, $b$-
$c$-, and $c$ - axes is shown for 1 bar, 2.5 kbar, and 5 kbar in
Fig. 3. Also shown is the case where stress was applied along an
arbitrary $b-c$ in-plane direction. The changes in $T_c$ for all
four experiments are shown in Fig. 4, and values of $\partial
T_{c}/\partial P_{a,b,c}$ from the present work are given in Table
1, which, by inspection, shows that the changes in $T_{c}$ are
substantially less (and/or non-monotonic) for in-plane stress,
when compared with either $a$-axis stress or hydrostatic pressure.

\subsection{Uniaxial stress dependence of the magnetoresistance at 0.5 K}
\subsubsection{Magnetoresistance for a-axis stress}
Further differences between inter-plane and in-plane stress on the
superconducting properties can be seen with magnetic field. In
Fig. 5 we present the magnetoresistance (MR)for $\kappa
$--(BEDT-TTF)$_{2}$Cu(SCN)$_{2}$ at 0.5 K for stress applied along
the a-axis. (Similar measurements for $a$-axis stress, which is
along the inter-plane direction, have been previously reported
\cite{campos95}, but have been re-measured in this work for
completeness). The data, which are offset for clarity, show
several distinct trends with increasing stress: the quantum
oscillation spectrum changes, the critical field $B_{c2}$
decreases, and the normal state resistance also decreases. The
Fourier spectrum for a-axis stress is shown in Fig. 6a. The
fundamental orbit $\alpha$ and its harmonic $2\alpha$ rapidly
attenuate with stress, but the magnetic breakdown orbit $\beta$
and its combination frequencies $\beta$ $\pm$ $\alpha$ present a
more complicated stress dependence (see Fig. 1b). At 2 kbar and
above, the magnetic breakdown orbit $\beta$ is dominant, and
$\alpha$ only appears in the $\beta + \alpha$ spectrum. In Fig. 6
b and 6c we show the a-axis stress dependence of the Fourier
spectrum. The main result of Fig. 6 is that the $\beta$ orbit,
which is proportional to the area of the first Brillouin zone (and
inversely proportional to the in-plane unit cell), decreases with
increasing a-axis stress. That the fundamental $\alpha$ increases
has been previously discussed by Campos et al.\cite{campos95}.
Although it is not possible to predict in detail, it is clear that
the symmetry of the Fermi surface, and therefore the underlying
unit cell, has been changed in some non-trivial manner. In Fig. 7
we show the decrease in the critical field with a-axis stress, as
determined from the maximum in the derivative dR/dB (see inset of
Fig. 7).
\subsubsection{Magnetoresistance for b-axis stress}
We selected the $b$-axis configuration for  high field studies
with in-plane stress since a change of lattice constant in this
direction would allow $t_{1}$ to increase with respect to $t_{2}$,
i.e. make the triangular lattice model more "equilateral". Also,
the $b$-axis strain showed the least decrease in $T_{c}$ of the
three principal axes (Fig. 4). The MR data for b-axis stress are
shown in Fig. 8 at 0.5 K. Unlike the a-axis results in Figs. 5-7 ,
the b-axis have a different Fourier SdH spectrum dependence on
stress, and the critical field $B_{c2}$ appears to increase with
stress. In Fig. 9a the Fourier spectrum shows that the magnetic
breakdown $\beta$ orbit amplitude more rapidly attenuates with
stress than does the $\alpha$ orbit amplitude. Also, the changes
in frequency are very small. To see them clearly, we show a single
SdH period in Fig. 9b, where a particular Landau level is observed
to move to lower fields with stress. This corresponds to a
decrease in the $\alpha$ orbit frequency, as shown in Fig. 9c. In
Fig. 10 we show the behavior of the critical field $B_{c2}$ as
defined from dR/dB (see inset of Fig. 10). The fact that $B_{c2}$
increases is the most dramatic effect of the in-plane stress in
the magnetoresistance.

\subsection{Uniaxial stress dependence of the magnetoresistance anomaly}

At higher temperatures, the critical field signature becomes
complicated by a magnetoresistance anomaly (defined as $\Delta R$
in Fig. 11) which appears in this materials, and which disappears
for in-plane magnetic field \cite{zoubrooks}. The higher
temperature behavior of this anomaly, which is a peak in the MR
just above $B_{c2}$, is shown for both stress directions in Fig.
11. For the a-axis case, with increasing stress, the peak rapidly
attenuates, and $B_{c2}$ decreases. For increasing b-axis stress,
the peak decreases, but $B_{c2}$ increases, as it did at 0.5 K
(see Fig. 10) for the highest stress values. In Fig. 12 we show
the full temperature and stress dependence of the anomaly from our
measurements as a function of $\Delta R$/R$_N$, where R$_N$ is
defined as the normal state resistance  extrapolated from fields
higher than $B_{c2}$ (see Fig. 11b).

\subsection{Analysis of normal state properties}

    Further information may be obtained from the normal state properties.
     From the temperature dependence of the SdH oscillation amplitudes the
      effective masses m* = m$_{SdH}$/m$_0$ of the carriers in the various
       orbits may be obtained, either from the wave forms directly, or from
       the amplitudes of the Fourier spectrum\cite{wosnitza}. A summary of our
       findings is shown in Fig. 13 for the high magnetic field data. For the
        a-axis data, the SdH effect could not be observed above 2.5 kbar ,
         although there is some evidence that m* is decreasing for the $\alpha$
         and $\beta$ orbits. Campos et al. also found that for a-axis stress,
          the SdH effect vanished at low values (of order 2 kbar), along
          with $T_c$ and $B_{c2}$. In the case of the b-axis , we could follow
          the SdH effect for the $\alpha$ orbit over the entire range of stress.
           Within the uncertainties of the determination, there is very little
            change in m*.
    In Figure 14 we present the stress dependent normal state resistance
     change for $a-$, $b-$, and $c-axis$ stress. The apparent
     difference in functional dependence between inter-plane and in-plane
      stress is represented by the two different fits (solid lines), as discussed in the next section.

\section{ Discussion}

\subsection{Uniaxial stress effects on crystallographic and electronic structure}
In order to gauge the effects of uniaxial stress on the
crystallographic structure, we consider the a-axis stress
dependence of the $\beta$ orbit, as shown in Fig. 6 and Table 2.
The frequency of the $F_\beta$ orbit is directly related to the
total area of the first Brillouin zone, $A_{BZ} = 2\pi/b *2\pi/c$.
For small changes, $\Delta F_{\beta}/F_{\beta}$ $\approx$ $\Delta
A_{BZ}/A_{BZ}$  $\approx$  -$\Delta$(bc)/bc. Hence the stress
dependence of $\beta$ is directly related to the in-plane lattice
parameters. We may therefore estimate an approximate linear
compressibility by considering the constant volume deformation of
a system with isotropic elasticity. For a cubic lattice of side
$a$, the fractional reduction in lattice constant (per kbar) along
one axis is $\epsilon = \delta a/a $ ($kbar^{-1}$). This will
produce an expansion (i.e. Poisson effect) of $\epsilon/2$ in the
orthogonal directions. Hence, using the cubic lattice-isotropic
compressibility approximation, $\Delta F_{\beta}/F_{\beta}$
$\approx$ $2\delta a/a$, and from Table 2, $\epsilon$ = -
0.0035/kbar for the a-axis compressibility of $\kappa
$--(BEDT-TTF)$_{2}$Cu(SCN)$_{2}$.

    From simple ``Golden rule" arguments, the conductivity due to
any single direction in the tight binding limit should increase as
$t^2$. Hence, for an approximately linear response of $t$ to
uniaxial stress along the a-axis, the a-axis conductivity change
would be $\sigma \approx \sigma_0 (1- 2\epsilon P_a)$. The
resistivity will therefore decrease inversely as $R(p)/R(0) = (1-
2\epsilon P_a)^{-1}$ with pressure. For the cases where both the
$a$-axis resistivity and stress are involved, this relation seems
to represent the data, with the caveat that the in-plane changes
in conductivity have been neglected. In reference to Fig. 14, we
find for the two $a$-axis cases studied, that $\epsilon$ =
$0.17\pm0.1$/kbar. The larger, fitted valued of $\epsilon$
(compared with the estimate above) from the normal state
resistance may reflect a stronger dependence of $t$ on $\epsilon$
than the linear approximation used in the argument above. In
contrast, the in-plane stress yields a very different pressure
dependence for the $a$-axis conductivity. Here, the decrease of
the normal state resistance can be fitted empirically by
$R(p)/R(0) = 1-(p/p_{c})^{2}$, where $p_{c}$ is $10.7\pm0.1$ kbar.
This strongly suggests that the effects of b-axis in-plane stress
on the transverse conductivity involve competing effects.

We next turn to stress induced changes in the Fermi surface
properties, which may be obtained by considering the SdH
oscillations. Returning to Fig. 6, we see two different
frequencies, $F_\alpha \approx 670 T$ and $F_\beta \approx 4200
T$. A summary of changes in $F_\alpha$ and $F_\beta $ for both
uniaxial and hydrostatic pressure is given in Table 2. In
reference to Fig. 1b, $F_\alpha$ arises from the closed orbit and
$F_\beta$ is a magnetic breakdown orbit which involves both the
closed and open orbit Fermi surface sections. The amplitude of
$F_\beta$ depends on the proximity of the closed and open orbit
Fermi surface sections, since a magnetic breakdown gap is involved
in the orbit. For the case of a-axis stress in Fig. 6, the shift
of relative amplitude from the $\alpha$ to $\beta$ Fourier
spectrum may indicate the closing of the magnetic breakdown gap.
Although the $\beta$ orbit decreases as expected for the expansion
of the real space b-c plane with a-axis stress, the $\alpha$
oribit increases significantly. A model for the sliding of the
(BEDT-TTF) molecules has been proposed to explain this
effect\cite{campos95,campos96}.
    In contrast, changes in the electronic structure for b-axis stress are very much reduced. Changes in $F_{\alpha}$ and $F_{\beta}$ are very small, and the $\alpha$ orbit is always dominant - hence the magnetic breakdown gap remains large. Since, for a reduction of the b-axis lattice constant, the c-axis lattice constant should increase by the Poisson argument, this may explain why $F_{\beta}$ does not change significantly. Although small, the b-axis change in $F_{\alpha}$ is opposite that for a-axis stress, and this indicates that stress in the two directions produce fundamental differences in the changes of electronic structure.

A treatment of the temperature dependence of the SdH amplitudes in
terms of the Lifshitz-Kosevich analysis\cite{wosnitza} yields
information about the effective mass m* (= m/m$_e$), which can be
enhanced by many body effects (electrons and phonons). We have
summarized our analysis of m* for a-axis and b-axis stress in Fig.
13., where the effective masses of the $\alpha$ and $\beta$ orbits
($m_{\alpha}$ and $m_{\beta}$)are shown as a function of stress.
$m_{\alpha}$ appears to be nearly stress-independent for both
axes, while $m_{\beta}$ begins to decrease for $P >$ 2 kbar.
%%%%%%%%%%%%%%%%%%%%%%%%%%%%%%%%%%
\subsection{Uniaxial stress effects on superconductivity}

In this final section we review the effects of uniaxial stress on
the superconducting state in relation to the observed changes in
the electronic structure, and comment on our results in the
context of recent theoretical work.
    One of the original motivating factors in this work was to use in-plane stress to change the ratio of $t_1$/$t_2$ as shown in Fig. 1a. Following arguments similar to those of the last section (for a cube of constant volume and isotropic elasticity), the application of stress along the b-axis will produce changes in $t_1$/$t_2$ of order (1 -$\epsilon P_b$)/(1 +$\epsilon P_b$/4), or for 10 kbar, about +4\%. (We note however, that since changes in $t$ are approximated as linear in $\epsilon$, the relative change in $t_1$/$t_2$ may be significantly larger, as the fit of $\epsilon$ in Fig. 14 indicates.)
For a-axis stress, a simple Poisson-like uniform expansion of the
in-plane lattice should not, to first order, cause the ratio
$t_1$/$t_2$ to change.

For comparison, Kondo and Moriya have used a two band Hubbard
model with spin fluctuations\cite{kondo98,kondo99} to compute the
dependence of of $T_{c}$ on $U/t_2$ and $t_{1}/t_{2}$. Referring
to Fig. 2 of Ref. \cite{kondo99}), these parameters are:
$t_{1}\approx 0.07 eV$; $t_{2}\approx 0.0875 eV$; and $U\approx
0.75 eV$. They predicted, using a d-wave symmetry for the order
parameter, that for fixed $t_{1}/t_{2}$ in the range 0.4 to 0.8,
$T_{c}/t_{2}$ versus $U/t_{2}$ would show a maximum change of
between +50\% and +10\%for intermediate values of ($U/t_{2}$)* in
the range 4 to 10. Vojta and Dagotto also have considered the
triangular lattice problem\cite{vojta} and find a tendency for
d-wave paring for $t_{1}/t_{2} \le 0. 8$. Clearly, we cannot
determine from our in-plane results accurate experimental changes
in $t_{1}/t_{2}$ and $U/t_{2}$, primarily due to the large
uncertainties in our estimates of $\epsilon$. Nevertheless our
in-plane data suggest that $T_{c}$, as well as the critical field
$B_{c2}$, follow  a significantly different dependence on in-plane
stress, when compared either with inter-plane stress, or with
hydrostatic pressure.

A second area of comparison with theory involves the magnetic
breakdown gap. Schmalian has used a two band Hubbard model with
spin fluctuations\cite{Schmalian} to predict d-wave
superconductivity for the $\kappa$-phase materials for the Fermi
surface described in Fig. 1. A key feature of this theory is the
inter-band coupling, which is influenced by the magnetic breakdown
gap region ("hot spots" in the terminology of Ref.
\cite{Schmalian}). Experimentally, we see that for $a-axis$ stress
this magnetic breakdown gap appears to be closing, and
correspondingly, the supeconducting transition and critical field
decrease. Following Fig. 3 of  Ref. \cite{Schmalian}, if the gaps
closed, then the additional nodes along the diagonals of the
d-wave gap symmetry would disappear, but this would not destroy
the d-wave state. Conversely, for $b-axis$ stress we note that the
magnetic breakdown gap does not close, the superconducting
transition is only a weak function of stress, and the critical
field actually increases.

A third point concerns the resistive anomaly which occurs in the
critical field behavior at temperatures of larger than 25 \% of
$T_c$, as shown in Figs. 11 and 12. This anomaly seems to be a
common feature of low dimensional, layered superconductors,
including high-$T_{c}$ cuprates and non-Cu based organic
superconductor \cite{su,zuo}. The effect is quite anisotropic, and
is found to be most significant when the current and the field are
applied perpendicular to the conducting plane\cite{zoubrooks}. The
reduction of $\Delta R$ with stress can be quantitatively
described by plotting $\Delta R$/$R_{N}$ versus stress as shown in
Fig. 12, where $\Delta R$ is the difference of resistance at $B$ =
$B_{\rm max}$ and $R_{N}$ is the normal state resistance obtained
from the extrapolation from $B$ $\gg$ $B_{c2}$ (See Fig. 11).
$\Delta R/R_{N}$ is largest at higher temperatures, and decreases
with stress for fixed temperature. This extra resistance has been
attributed to various mechanisms. One involves the presence of
magnetic impurities \cite{mielke}, but since the impurity
concentration would not change with stress, the present work is
inconsistent with a magnetic impurity effect. There are several
other models which treat this effect in terms of interplaner
coupling effects. One involves the inter-layer resistance due to
Josephson junction coupling\cite{zuo}. A second model is that of
Maki and co-workers\cite{maki}. Here the anomaly is treated as a
superconducting fluctuation mechanism, with the conclusion that
the interplane magnetoresistance can only arise from d-wave
symmetry. Finally, Kartsovnik and co-workers\cite{kartsovnik}
conclude that the interlayer transport in the superconducting
state is mediated by superconducing-normal-superconducting
Josephson tunneling. Both models support the highly ansiotropic
nature of the anomaly in tilted fields. Since $\Delta R$ is most
sensitive to the $a-axis$ interplaner stress, interplaner coupling
effects may indeed be the origin of this effect.

Finally, we note dependence of the effective masses on uniaxial
stress shown in Fig. 13.  If we assume the change of $T_{c}$
follows the weak-coupling BCS scheme, $T_{c}$ is a strong function
of effective mass $m^{*}$. This trend does appear to some extent
in the a-axis data, with both $m^*$ (over the small range
measured) and $T_{c}$ decreasing. For the b-axis data, the weak
dependence of $m^*$ and $T_{c}$ on stress appears to be mutually
consistent.

\section{Conclusions}

In summary, we have employed both inter-planar (transverse) and
in-plane uniaxial stress to physically change parameters
associated with the triangular lattice model for superconductivity
in the quasi-two dimensional metal $\kappa
$--(BEDT-TTF)$_{2}$Cu(SCN)$_{2}$. Our main results show that the
superconductivity associated with inter-planar stress is rapidly
reduced, as is observed in the critical temperature, the critical
field, and also in the resistive anomaly associated with the
critical field. The normal state resistance follows a simple model
for the increase of inter-planar band-width with stress. For
in-plane stress, a more direct modification of the parameters of
the triangular lattice model is possible. Indeed, for in-plane
stress, the critical temperature is a considerably weaker function
of stress, and may even increase slightly over a small range.
Also, the critical field is observed to increase significantly,
and the resistive anomaly associated with the critical field is
not a strong function of in-plane stress. Here the normal state
resistance is a more complicated function of in-plane stress. We
have used the SdH effect to monitor stress-induced changes in the
unit cell and electronic structure. The most dramatic changes
occur for the a-axis case where the fundamental closed orbit area
increases by 10\%, and the inter-band magnetic breakdown gap
essentially disappears. In contrast, the in-plane stress does not
significantly change the SdH parameters associated with the Fermi
surface, except that the fundamental closed orbit decreases with
b-axis stress. Also, the magnetic breakdown gap remains open.

    We find provocative, but inconclusive comparisons with contemporary theories for d-wave superconductivity in this "kappa phase" material. Over the
 the range of parameters investigated,
there is no single aspect of the electronic structure to which the
superconductivity is highly sensitive, but we emphasize that the
in-plane modification of the lattice parameters has the most
unusual effects on the superconducting state. Application of
strictly uniaxial stress via methods now employed by several
groups \cite{kagoshima} \cite{murata}, coupled by higher values of
stress, perhaps to 20 kbar, should allow  the ratio $t_1$/$t_2$ to
change enough to more fully test the predictions of the triangular
lattice Hubbard models.

\bigskip
\noindent {\em Acknowledgements} This work was supported in part
by NSF-DMR95-10427 and DMR99-71474. The work was carried out at
the National High Magnetic Field Laboratory, supported by a
contractual agreement between the State of Florida and the NSF
through NSF-DMR-95-27035.
%%%%%%%%%%%%%%%%%%%%%%%%%%%%%%%%%%%%%%%%%%%%%%%%%%%%%%%%%%%%%%%%%%%

%%%%%%%%%%%%%%%%%%%%%%%%%%
\newpage
%%%%%%%%%%%%%%%%%%%%%%%

\begin{center}
\begin{table}
\begin{tabular}{|l|c|c|c|} \hline
  % after \\ : \hline or \cline{col1-col2} \cline{col3-col4} ...
\multicolumn{4}{|c|} {Uniaxial and Hydrostatic Pressure Dependence
of} \\ \multicolumn{4}{|c|} {$T_{c}$ for
$\kappa$(ET)$_{2}$Cu(NCS)$_{2}$: $\Delta T_{c}/\Delta P$ (K/kbar)}
\\ \hline
& $a$-axis  & $b$-axis & $c$-axis
\\ \hline
 Uniaxial (This work samples 1,2 4, and 5  & -0.78  & $0+ <$ 3 kbar  &-0.2 \\
   &&$-0.2<3$ kbar& \\ \hline
 Uniaxial (This work-sample 3)&--&\multicolumn{2}{|c|}{$+0.3<2$ kbar} \\
 &&\multicolumn{2}{|c|}{$-0.08>2$ kbar}\\ \hline
 Uniaxial\cite{campos95}&-2&--&--\\ \hline
Tensile stress ($\Delta T_c$) \cite{kusuhara}&--&+0.5 $\sim$
0.8K&--\\ \hline Thermal expansion
(Kund)\cite{kund}&-3.2&0.0&+1.46 \\ \hline Thermal expansion
(Lang)\cite{lang}&-6.2&-0.14&+3.44 \\ \hline
\multicolumn{1}{|l|}{Hydrostatic\cite{caulfield}}&
\multicolumn{3}{|c|}{-3} \\ \hline
\end{tabular}
\vspace{0.5 cm}\caption{Summary of the effects of uniaxial and
hydrostatic pressure on the superconducting transition temperature
$T_c$ of $\kappa $--(BEDT-TTF)$_{2}$Cu(SCN)$_{2}$. } \vspace {2.0
cm }
\end{table}
\end{center}

\begin{center}
\begin{table}
\begin{tabular}{|c|c|c|} \hline
\multicolumn{3}{|c|} {Uniaxial and Hydrostatic Pressure Dependence
of SdH Frequencies:}\\ \multicolumn{3}{|c|} {$\kappa
$--(BEDT-TTF)$_{2}$Cu(SCN)$_{2}$: $\Delta F/F_{0}/\Delta P$ (\%
kbar)} \\ \hline Uniaxial (this work)&$F_{\alpha}$&$F_{\beta}$\\
\hline $a$-axis&+3.0\cite{campos95}, 5.0&-0.7\\ \hline $b$-axis
&-0.25&+0.1\\\hline Hydrostatic\cite{caulfield} &+1.84&+0.36\\
\hline
\end{tabular}
\vspace{0.5 cm} \caption{Summary of the effects of uniaxial and
hydrostatic pressure on the Fermi Surface topology of
$\kappa$--(BEDT-TTF)$_{2}$Cu(SCN)$_{2}$.}
\end{table}
\end{center}

\newpage

\begin{figure}[htbp]
\caption{a) Molecular stacking (the oval structure indicates each
BEDT-TTF molecule) and the equivalent triangular lattice with
transfer integrals $t_1$ and $t_2$. b) The Fermi surface of
$\kappa $--(BEDT-TTF)$_{2}$Cu(SCN)$_{2}$ at ambient pressure. The
fundamental $\alpha$ and breakdown $\beta$ orbits are also shown.}
\end{figure}

\begin{figure}[htbp]
\caption{Resistance vs. temperature as a function of applied
uniaxial stress along the c-axis for Sample 1. Inset: $T_{c}$
values are obtained from gaussian fits to the peaks in the
derivative dR/dT.}
\end{figure}

\begin{figure} [htbp]
\caption{Summary of resistance vs. temperature data for 0, 2.5,
and 5 kbar for all uniaxial stress directions.}
\end{figure}

\begin{figure} [htbp]
\caption{Summary of the uniaxial stress dependence of the
superconducting transition $T_{c}$ for all stress directions
studied.}
\end{figure}

\begin{figure}[htbp]
\caption{Magnetoresistance of sample 5 at 0.5 K for different
values of uniaxial stress along the a-axis.}
\end{figure}

\begin{figure}[htbp]
\caption{a) Fourier spectrum of sample 5 at 0.5 K for different
values of uniaxial stress along the a-axis. b) a-axis stress
dependence of the $\beta$ orbit and its difference frequencies. c)
a-axis stress dependence of the $\alpha$ orbit. Above 1 kbar,
$F_\alpha$ was determined from the difference frequencies in b).}
\end{figure}

\begin{figure} [htbp]
\caption{Critical field $B_{c2}$ of sample 5 at 0.5 K vs. a-axis
stress. Inset: The peaks in the derivatives dR/dB were used to
determine the $B_{c2}$ values.}
\end{figure}

\begin{figure}[htbp]
\caption{Magnetoresistance of sample 4 at 0.5 K for different
values of uniaxial stress along the b-axis.}
\end{figure}

\begin{figure}[htbp]
\caption{a) Fourier spectrum of sample 4 at 0.5 K for different
values of uniaxial stress along the b-axis. b) Shift of the
$\alpha$ orbit SdH waveform with b-axis stress. c) b-axis stress
dependence of the $\alpha$ orbit derived from b).}
\end{figure}

\begin{figure} [htbp]
\caption{Critical field $B_{c2}$ of sample 4 at 0.5 K vs. b-axis
stress. Inset: The peaks in the derivatives dR/dB were used to
determine the $B_{c2}$ values.}
\end{figure}

\begin{figure} [htbp]
\caption{a) Resistive anomaly at 2.0 K in the MR of sample 5 for
stress applied along $a$-axis. b) Resistive anomaly at 2.0 K in
the MR of sample 4 for stress applied along $b$-axis. Note that
the resistive anomaly is defined as the difference $\Delta R$
divided by the extrapolated normal state resistance $R_N$ shown as
the dashed line. A similar construction was used to determine
$\Delta R$/$R_N$ for all data shown.}
\end{figure}

\begin{figure}[htbp]
\caption{a) Resistive anomaly $\Delta R$/$R_N$ for a-axis stress
at different temperatures. $\Delta R$/$R_N$ for b-axis stress at
different temperatures.}
\end{figure}

\begin{figure}[htbp]
\caption { Normal state (fractional) resistance change vs. applied
stress for samples 4 and 5 (at 26 T and 0.5 K) and for .samples 1
and 2 were obtained at (at 0 T and 14 K ). The solid lines are
fits to the $1-(P/10.7)^{2}$ (sample 4) and $(1+0.34P)^{-1}$
(sample 2 and 5), where $P$ is given in kbar.}
\end{figure}

\begin{figure}[htbp]
\caption{Effective masses for $\alpha$ and $\beta$ orbits vs.
uniaxial stress.}
\end{figure}

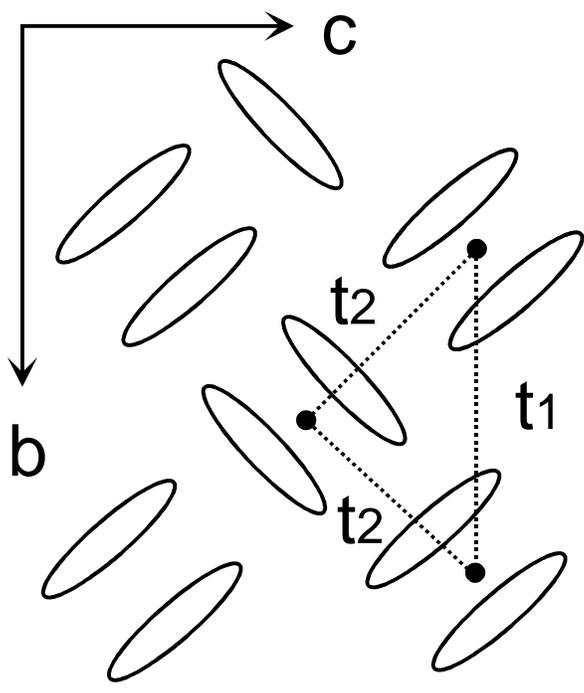
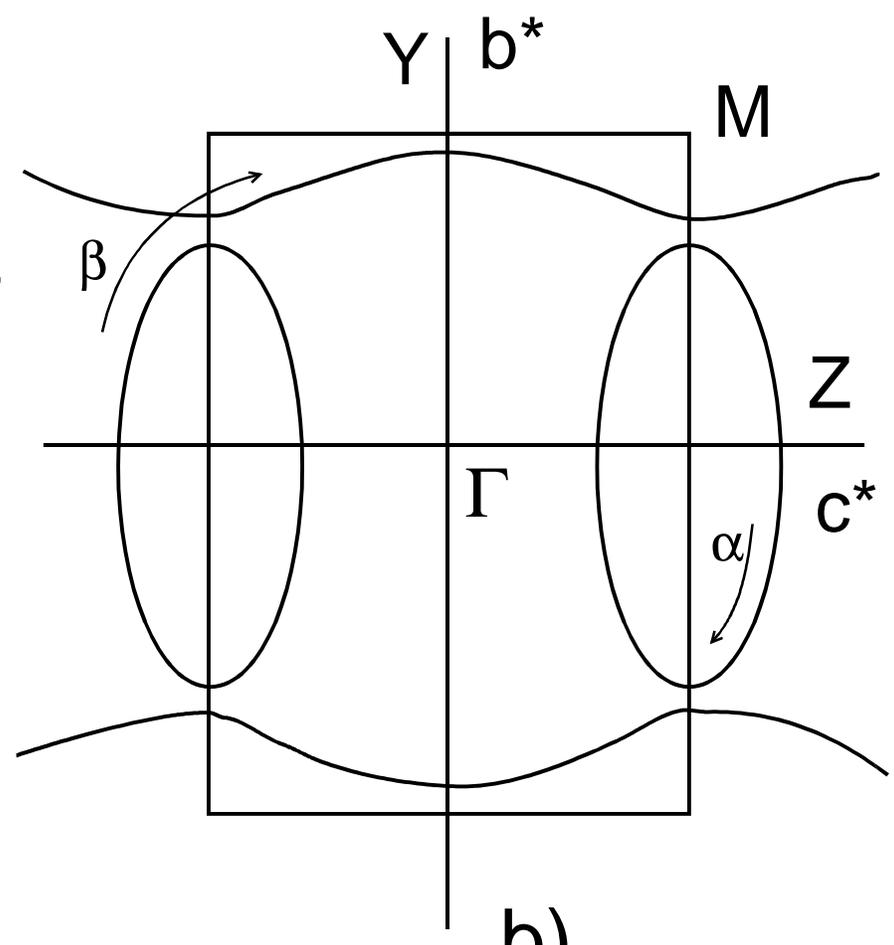
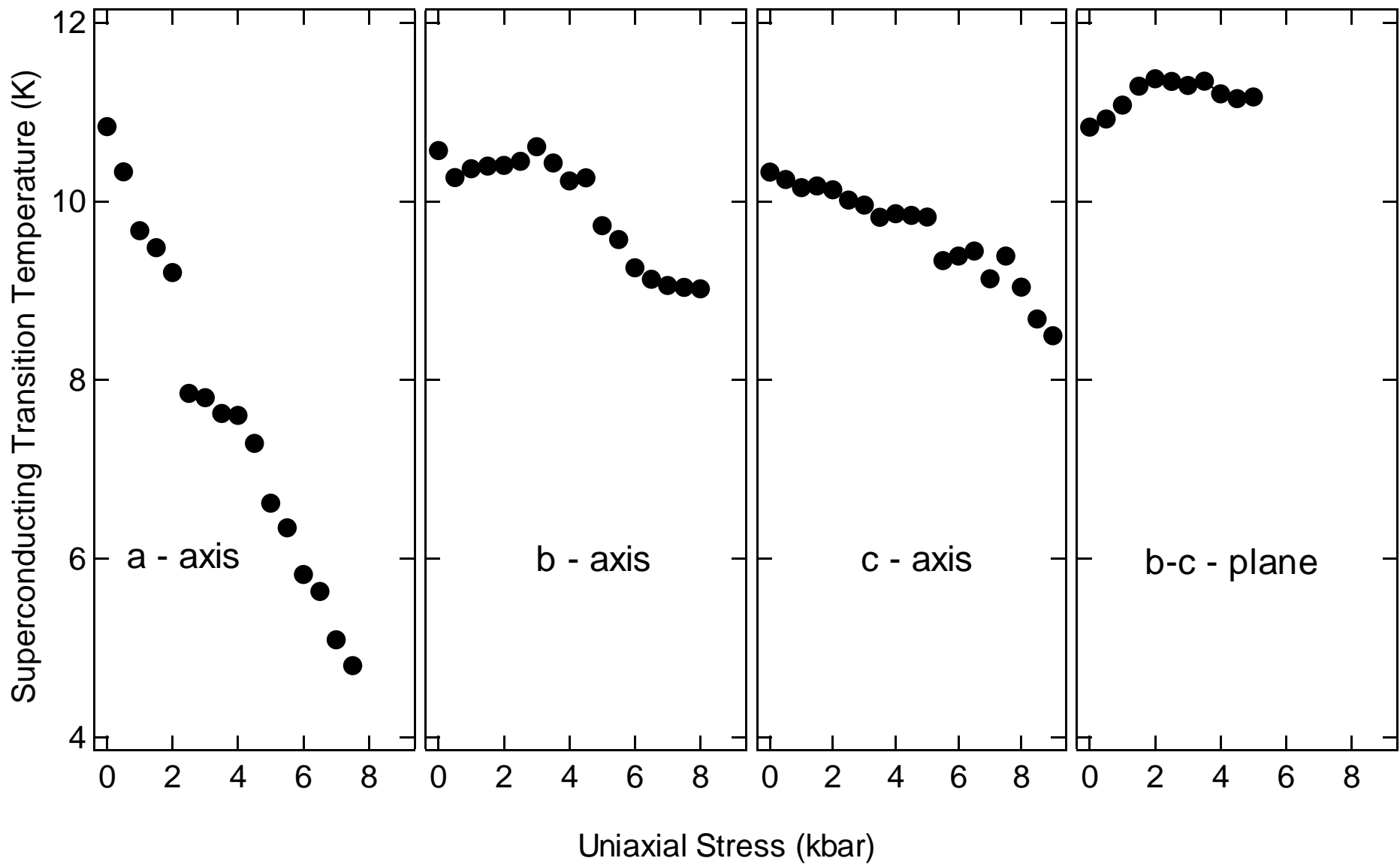
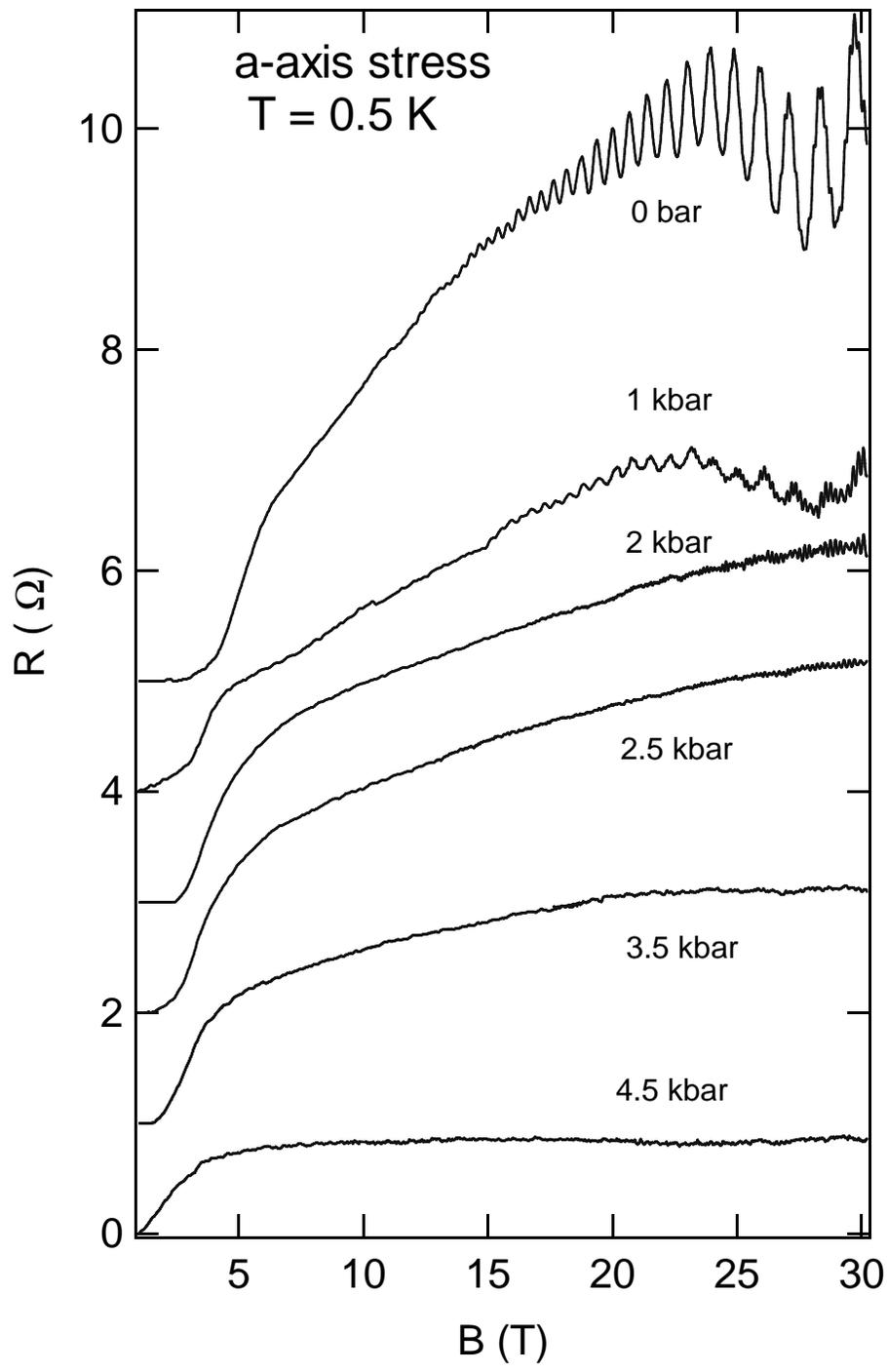
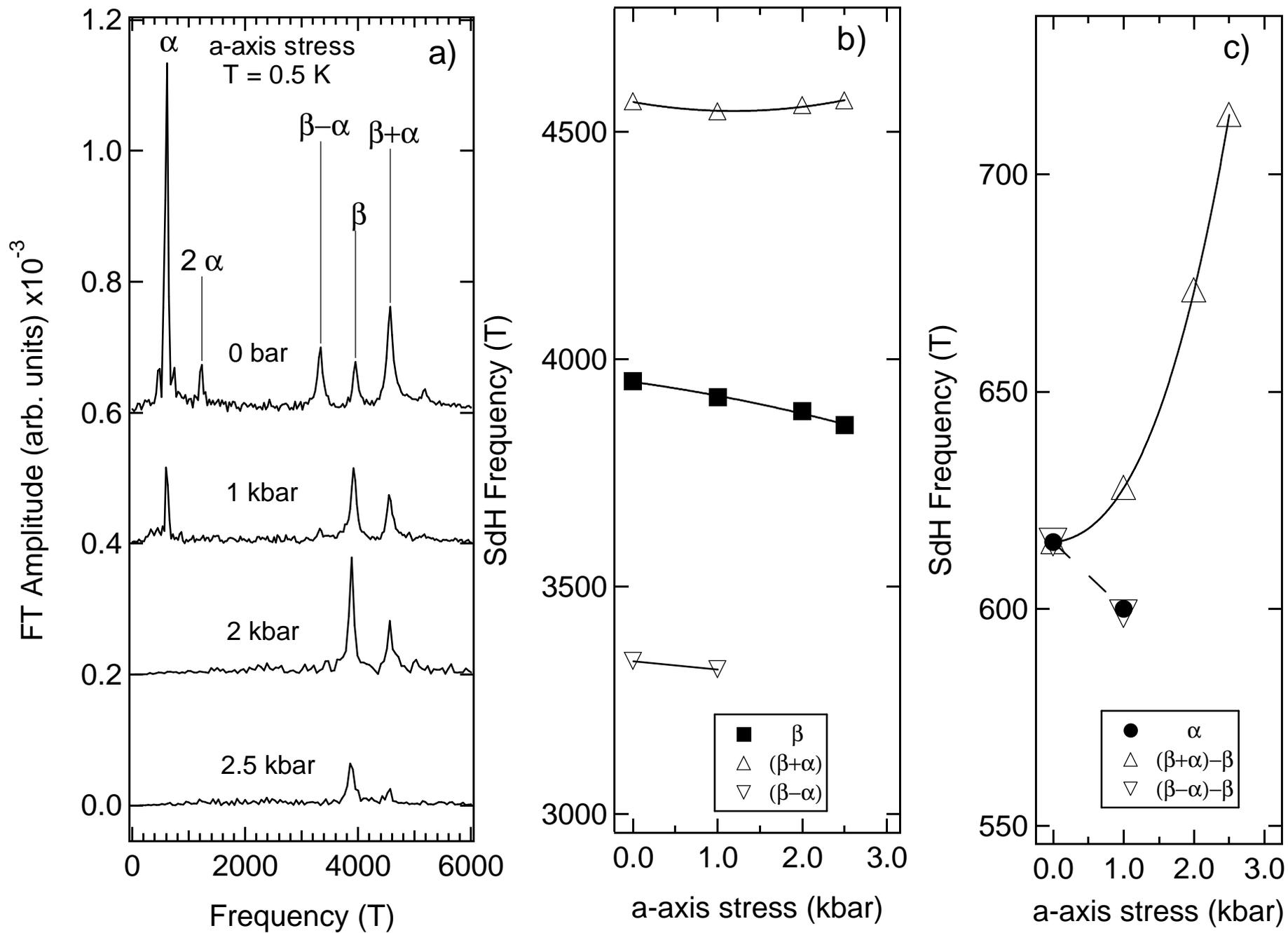
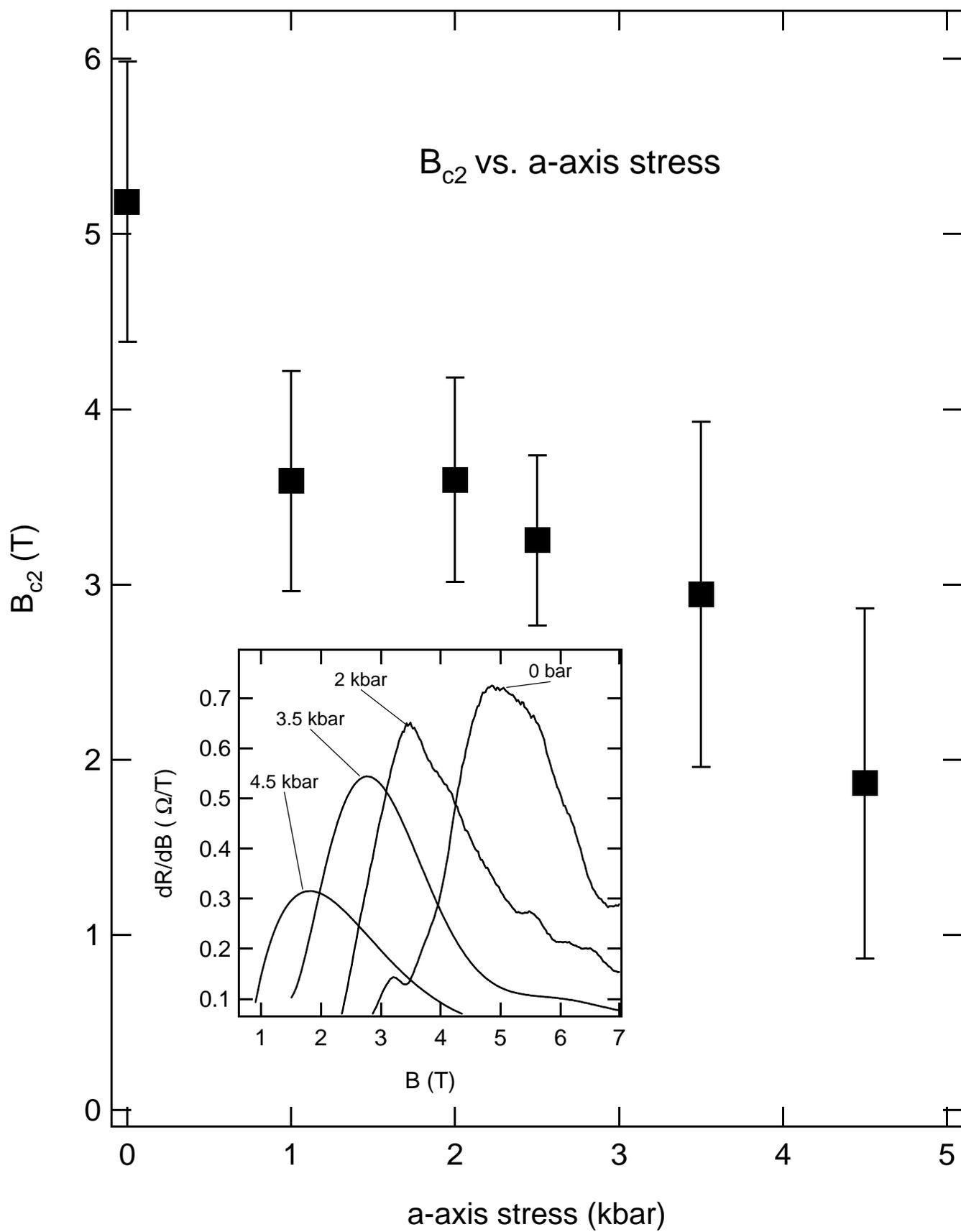
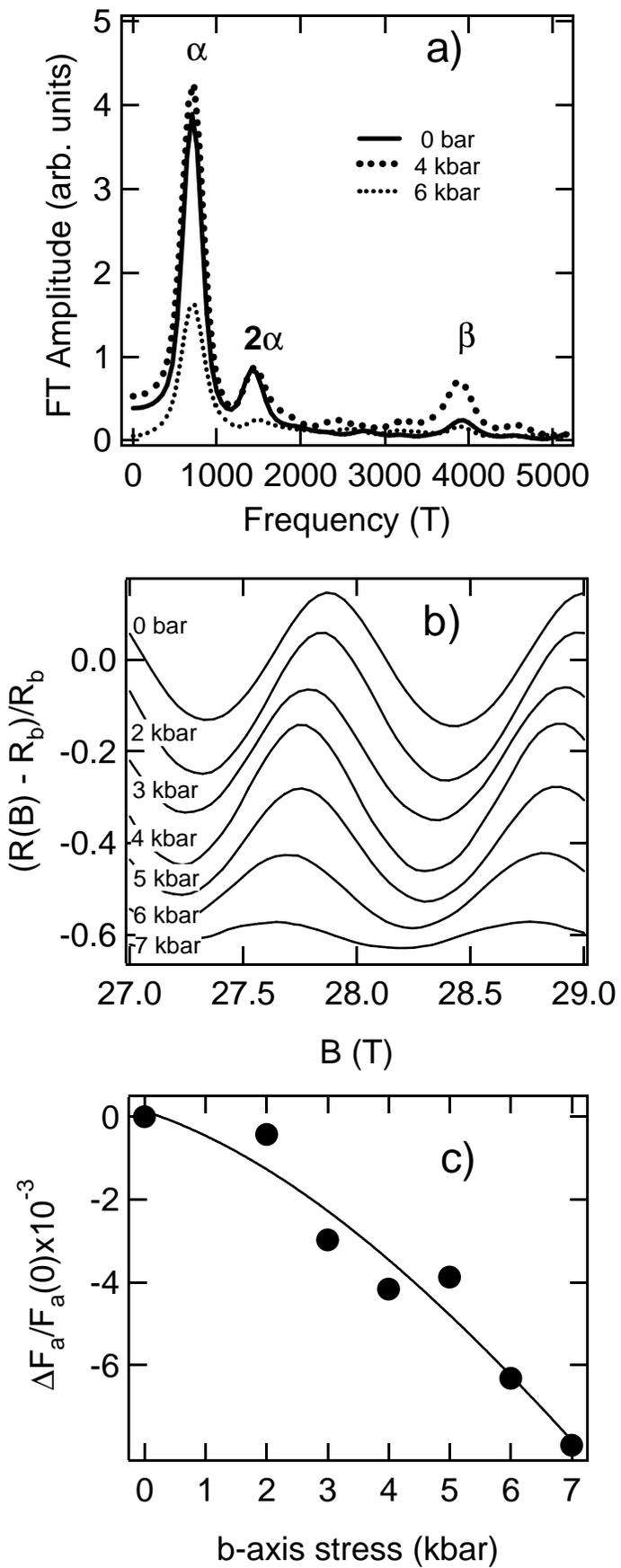
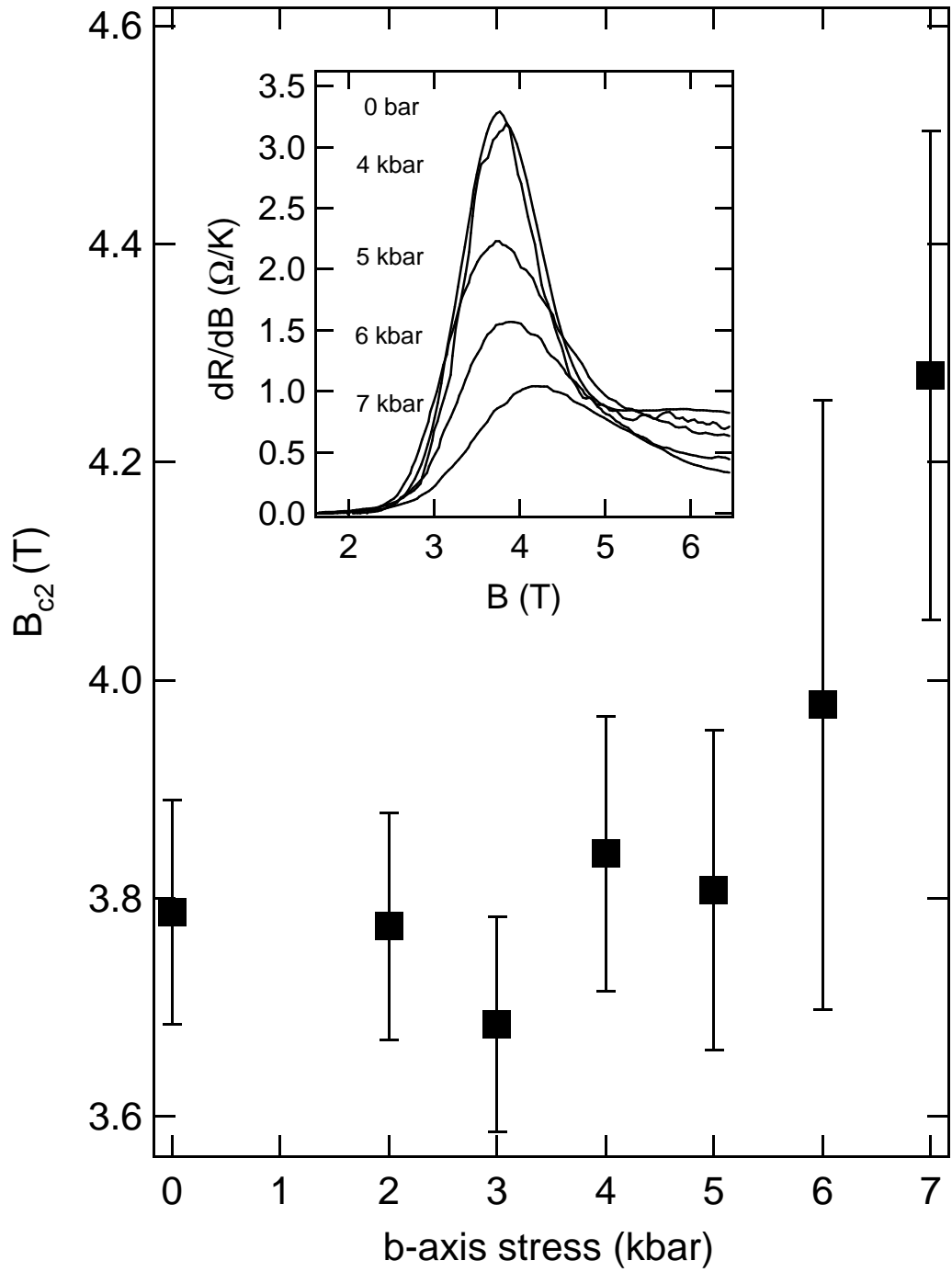
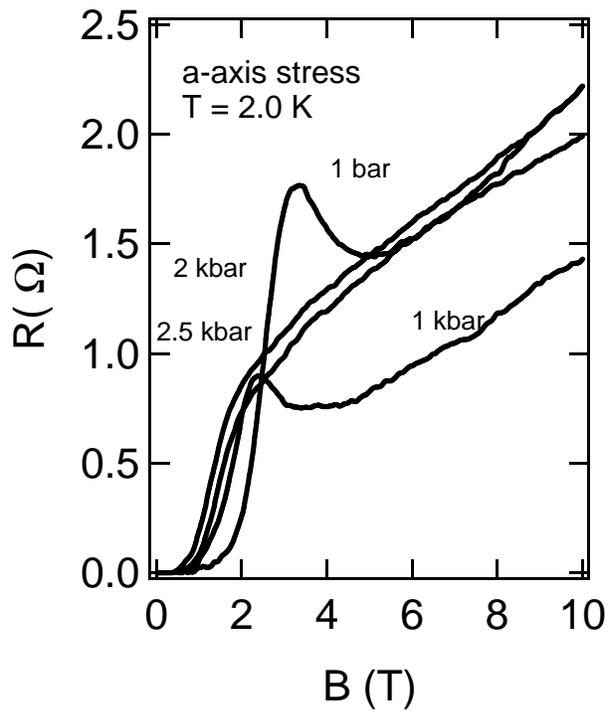
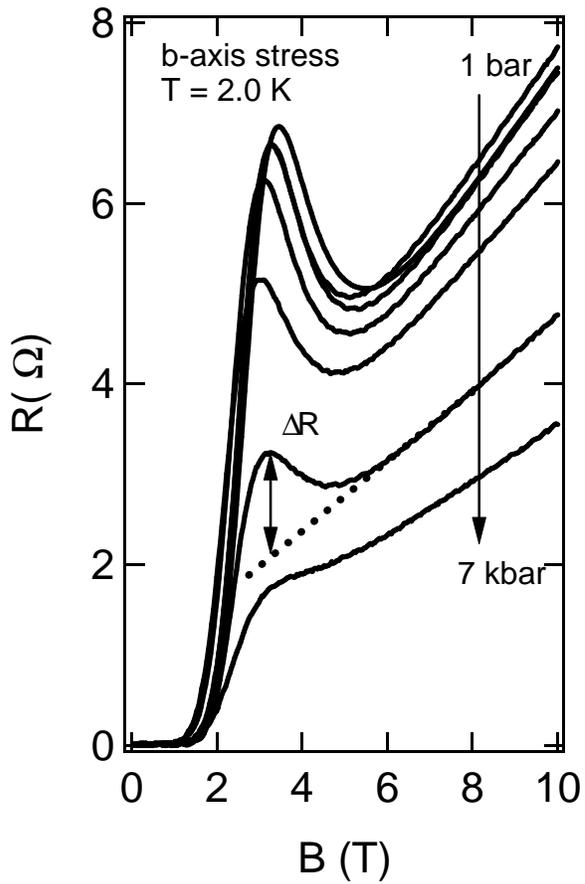
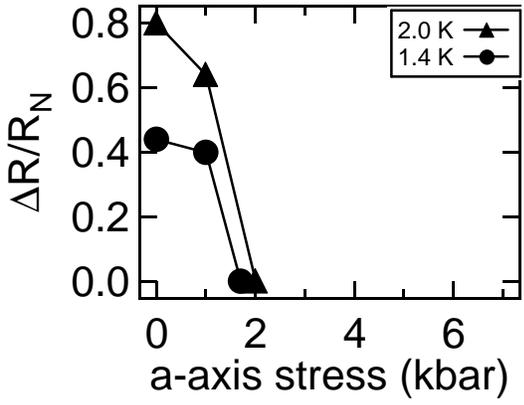
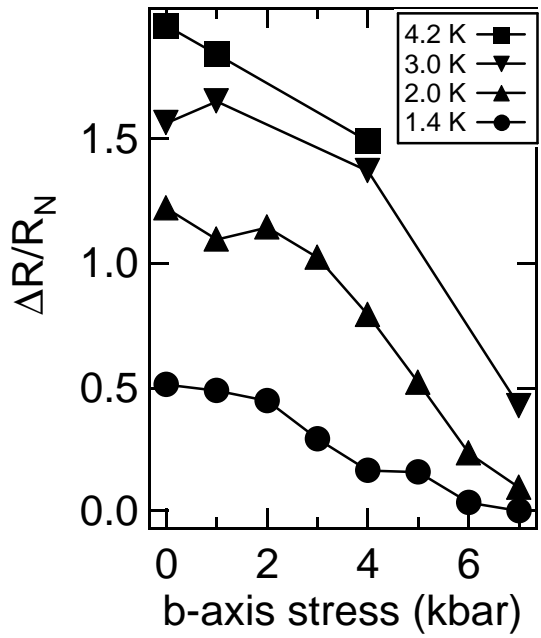
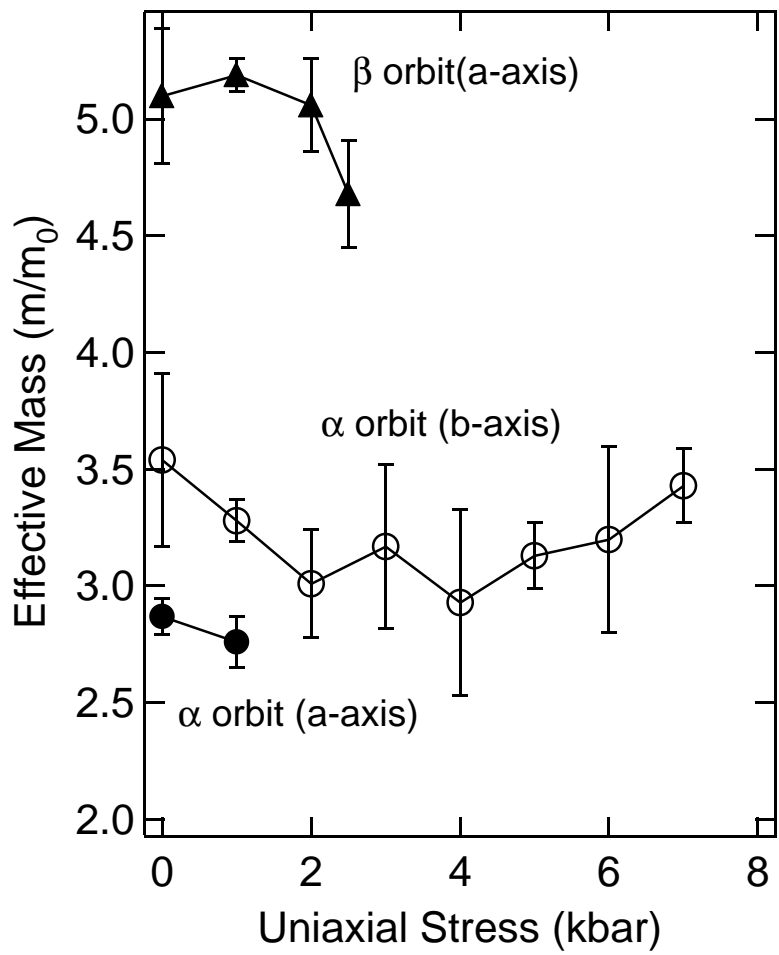
\end{document}